# Electromagnetic-induced fission of $^{238}$U projectile fragments, a test case for the production of spherical super-heavy nuclei

A. Heinz[1,a], K.-H. Schmidt[1], A. R. Junghans[2,b], P. Armbruster[1], J. Benlliure[1,3], C. Böckstiegel[2], H.-G. Clerc[2], A. Grewe[2], M. de Jong[2], J. Müller[2], M. Pfützner[4], S. Steinhäuser[2], B. Voss[1]

[1] Gesellschaft für Schwerionenforschung, Planckstr. 1, 64291 Darmstadt, Germany
[2] Institut für Kernphysik, Technische Universität Darmstadt, Schloßgartenstr. 9, 64289 Darmstadt, Germany
[3] Faculdad de Fisica, Universidad de Santiago de Compostela, 15706 Santiago de Compostela, Spain
[4] Instytut Fizyki Doswiadczalney, Uniwersytet Warszawaski, ul Hoza 69, 00-381 Warszawa, Poland

**Abstract:** Isotopic series of 58 neutron-deficient secondary projectiles ($^{205,206}$At, $^{205-209}$Rn, $^{208-212,217,218}$Fr, $^{211-223}$Ra, $^{215-226}$Ac, $^{221-229}$Th, $^{226-231}$Pa, $^{231-234}$U) were produced by projectile fragmentation using a 1 $A$ GeV $^{238}$U beam. Cross sections of fission induced by nuclear and electromagnetic interactions in a secondary lead target were measured. They were found to vary smoothly as a function of proton and neutron number of the fissioning system, also for nuclei with large ground-state shell effects near the 126-neutron shell. No stabilization against fission was observed for these nuclei at low excitation energies. Consequences for the expectations on the production cross sections of super-heavy nuclei are discussed.

## 1. Introduction

The influence of the nuclear shell structure on fission has been established early after the discovery of the nuclear fission process itself and is responsible for phenomena as different as the observed asymmetric fission fragment mass distributions (see e.g. [1]) and the stability of the heaviest

---

[a] Present address: Argonne National Laboratory, Physics Division, Bd. 203, 9700 South Cass Avenue, Argonne, Il 60439, USA
[b] Present address: Nuclear Physics Laboratory, BOX 354290, University of Washington, Seattle, Wa 98195, USA

observed nuclei with respect to spontaneous fission [2]. Indeed, it is the complex interplay of microscopic and macroscopic effects, which turns nuclear fission into the fascinating and puzzling process it is. The present work focuses on the question, how a pronounced shell structure of a heavy fissile nucleus affects its de-excitation and its survival probability against fission.

The half-lives and ground-state-decay properties of the heaviest known nuclei are essentially determined by shell structure. In particular, spontaneous-fission half-lives are extremely sensitive to the magnitude of the ground-state shell effect [2,3]. However, it is known that this stabilization against fission vanishes with excitation energies well above the fission barrier. It is an experimental challenge to determine how an increase in excitation energy influences the shell structure of a nucleus and thereby the competition between fission and the other decay modes of the excited compound nucleus.

This is also a crucial question for a deeper understanding of the production of super-heavy nuclei [4]. Unfortunately, their reaction rates are far too low to systematically investigate their formation mechanism. The heaviest nuclei unambigiously identified are predicted to be strongly deformed in their ground state [5]. First attempts to produce spherical super-heavy nuclei near the next major neutron and proton shells above $^{208}$Pb have been made [6, 7]. But as the observed decay chains could not be linked to known alpha decays the production of spherical super-heavy elements with Z=114,116 needs to be confirmed. Excitation functions of the formation cross sections of spherical super-heavy nuclei are missing.

The heaviest known doubly magic nucleus, which is accessible to experimental investigation, is $^{208}$Pb. However, this nucleus is highly stabilized against fission due to its macroscopic properties alone, which makes it extremely difficult to observe fission at sufficiently low excitation energies above the fission barrier. Therefore, we chose to investigate radioactive proton-rich nuclei in the vicinity of the 126-neutron shell. Those nuclei have already been studied before in a similar context. In a first series of experiments [8], these nuclei have been produced with rather high excitation energies (a few tens of MeV) and high angular momenta using fusion-evaporation reactions. Evaporation-residue cross sections have been measured. No evidence for the suppression of fission in the vicinity of the 126-neutron shell was observed. In a more recent experiment [9], the production of heavy proton-rich nuclei after projectile fragmentation of relativistic $^{238}$U has been studied. This experiment produced nuclei around $N$=126 with lower angular momenta [10], but still did not give an indication of an enhanced survival probability with respect to fission. This finding has been attributed to the influence of collective excitations on the level density. The fission decay probability depends on the level density above the fission barrier, normalized by the level density of the daughter nucleus produced by neutron evaporation above the ground state. If the daughter nucleus is spherical, its excited levels consist only of single-particle and vibrational excitations, while the level density above the fission barrier is enhanced due to additional rotational excitations. This leads to an increased fission probability. Thus, the collective enhancement counteracts the stabilisation against fission by the ground-state shell effect in magic nuclei [8, 9].

The present work forms the continuation of a previous systematic study on the conditions for the synthesis of heavy elements [8]. While the influence of nuclear structure on the entrance channel was comprehensively studied and clearly demonstrated, the fission competition in the deexcitation process did not exhibit the expected stabilization. The advanced technical installations of GSI allow us now to revisit this problem with a new experimental approach. Here, we present an experimental study of fission of relativistic secondary projectiles after electromagnetic interactions. This technique represents considerable progress, since it allows measurements of fission cross sections at low angular momenta and at low excitation energies close to the height of the fission barrier. The demand for such a study, has been emphasised recently [11].

Our experimental approach as well as the physics of electromagnetic-induced fission has already been described in a previous publication [12] in great detail. But while this publication concentrated mainly on fission-fragment charge distributions and their interpretation, the present work will focus on the measurement of low-energy fission cross sections.

## 2. Experimental set-up

The experimental results discussed in this article were obtained at the secondary-beam facility of Gesellschaft für Schwerionenforschung (GSI). The heavy-ion synchrotron SIS delivered a primary beam of $^{238}$U at an energy of 1 $A$ GeV, with an average intensity of $10^7$ ions per second. The beam impinged on a 657 mg/cm$^2$ beryllium target, which was located at the entrance of the fragment separator (FRS) [13]. At the given primary-beam energy, a large number of mostly proton-rich isotopes is produced in peripheral collisions via relativistic projectile fragmentation [9]. The fragment separator, with its ability to spatially separate and identify projectile fragments event-by-event, was used to prepare beams of 58 nuclides between $^{205}$At and $^{234}$U (see Figure 1) whose fission cross sections after nuclear and electromagnetic interaction in a secondary lead target were measured in a dedicated detector set-up. With this experimental approach, it was even possible to investigate short-lived nuclei, such as $^{216}$Ra and $^{217}$Ac. With their half-lives in the order of 100 ns, about half of the nuclei produced in the beryllium target reach the exit of the fragment separator. In the following, the preparation of the secondary beams as well as the measurement of the fission cross sections will be described.

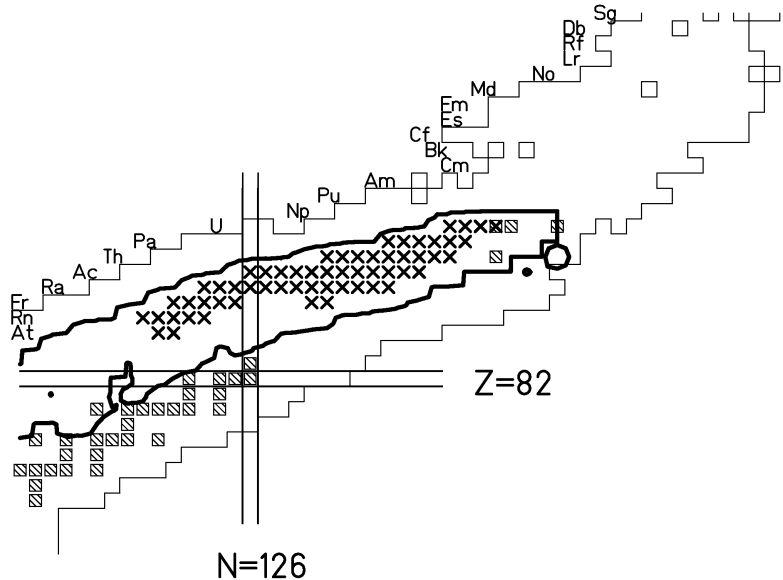

Figure 1: Chart of the nuclides. The area in which the measured production cross sections for projectile fragments from the $^{238}$U at 950 $A$ MeV on copper reaction are larger than 0.1 mb is marked by a boundary line [9]. That measurement [9] did not include all nuclei in the area, which causes the irregularities of the boundary in this figure. Nuclei investigated in the present work are indicated (x).

A schematic drawing of the fragment separator with the detector system used is shown in Figure 2. At the entrance of the fragment separator, a secondary-electron transmission monitor (SEETRAM) is located. It is used to measure the primary-beam intensity [14]. The fragment separator deflects the reaction products according to their mass-to-charge ratio in its first two dipoles. An aluminium degrader, located at the intermediate focal plane of the separator, is followed by another two magnetic dipoles. In the experiment discussed here, the fragment separator was used in its achromatic mode [15]. The degrader thickness was chosen to be about 50 % of the range of the projectile fragments, which was about 3.5 g/cm$^2$ with slight variations depending on the selected fragments.

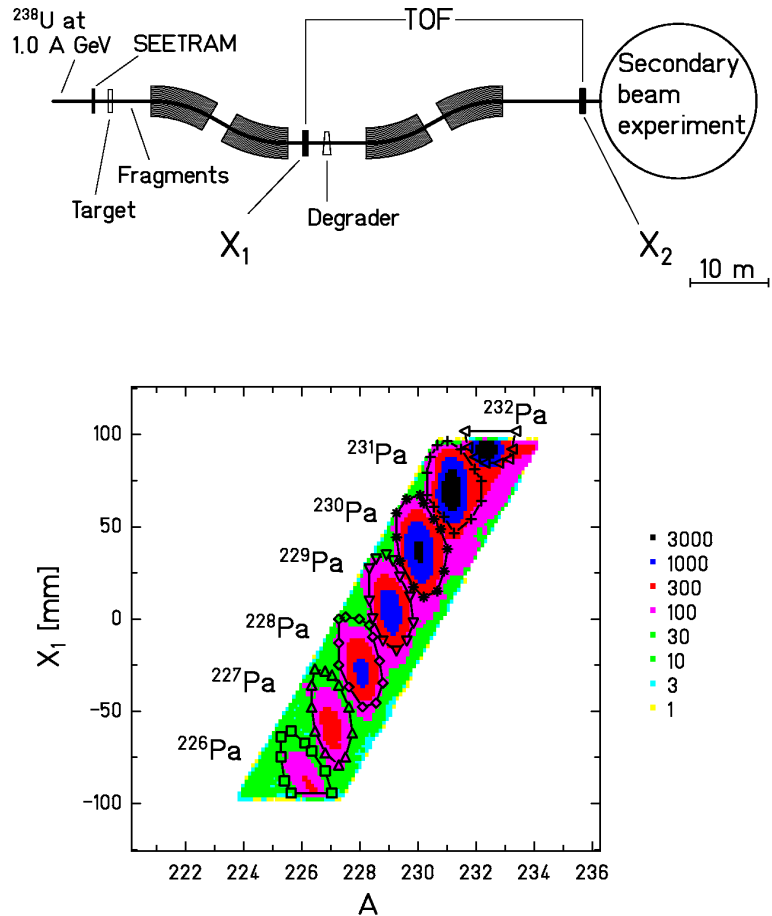

Figure 2: Top: Schematic drawing of the fragment separator as it was used in the experiment described here. Bottom: Identification spectrum using a chain of protactinium isotopes as an example. Plotted is the position at the central focal plane as function of the nuclear mass of the secondary beam. The scale indicates the number of counts per channel. The conditions which were used in the analysis for the individual isotopes are indicated.

The horizontal positions of the fragments at the central and at the final focal plane were determined, using position-sensitive plastic scintillation detectors. The time-of-flight between both detectors was measured as well. In order to determine the angle of the projectile fragments at the exit of the fragment separator with respect to the centred beam, two multi-wire proportional counters, not shown in Figure 2, were installed. This detector set-up, described in detail in reference [12], is sufficient to identify the projectile fragments according to their nuclear charge and mass on an event-by-event base. As the secondary beams of interest here had a rather high nuclear charge, different ionic charge states of the projectile fragments might cause ambiguities in the identification procedure. A layer of 212 mg/cm$^2$ niobium downstream from the target and a second foil of 105 mg/cm$^2$ niobium behind the degrader were mounted in order to maximize the amount of fully stripped ions behind the target and behind the degrader, respectively. A complete list of the different layers of matter in the beam-line with their thicknesses is given in reference [12].

The set-up to measure fission cross sections of secondary beams at the exit of the fragment separator is shown in Figure 3. The first detector is the position-sensitive scintillation detector, located at the final focal plane of the fragment separator, which was shown already in Figure 2. It was used for the identification of the projectile fragments, as described above, but it is as crucial for the measurement of the fission cross sections, as will be described in the following section. It also served as a start detector to measure the time-of-flight of the fission fragments.

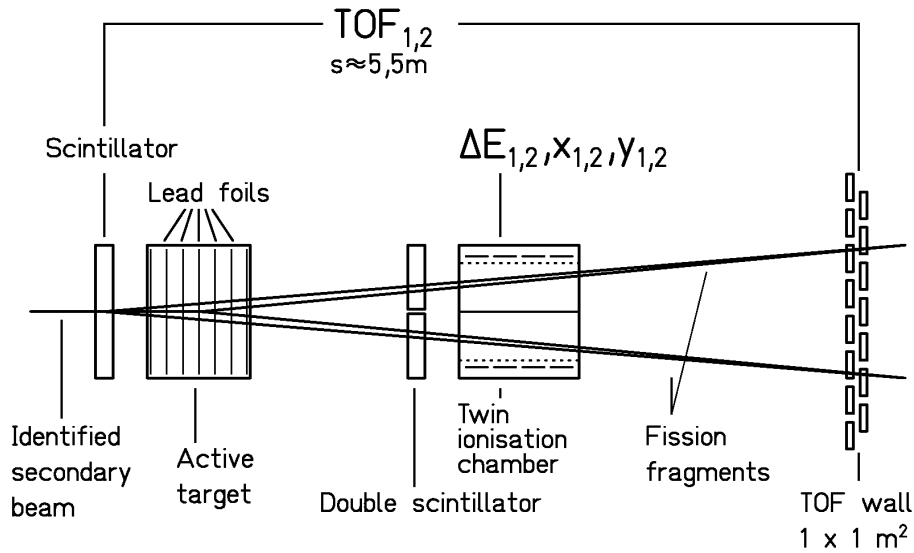

Figure 3: Schematic drawing of the experimental set-up at the final focal plane. This set-up was optimised to detect in-flight fission of relativistic secondary beams after electromagnetic interaction.

The second detector is the so-called active target. Five lead foils with a total thickness of 3.03 g/cm$^2$ are mounted inside a gas-filled detector chamber with a 0.027 g/cm$^2$ aluminium foil before and behind the target foils, respectively. By applying appropriate voltages, the active target acts as subdivided ionisation chamber for detecting a change in the energy loss of the traversing ions. As a fission event reduces the total energy loss by about a factor of two with respect to the incoming projectile fragment, this detector determined the target foil in which fission took place, and it discriminated fission events occurring before or after the lead target foils.

Downstream from the active target, two plastic scintillation detectors were located, mounted on top of each other. They were used to provide a fast trigger for fission events and for normalization purposes. These detectors selected fission events by putting a condition on the event multiplicity, which was used for the fast trigger. The difference in energy loss of fission fragments and secondary-beam particles was utilized for a more precise selection of fission events in the data analysis. The efficiency for the detection of fission events by the scintillation detectors was determined by a Monte-Carlo simulation to be 90 %

The next detector was a large twin ionisation chamber. Two active volumes shared one common cathode. The anodes were subdivided into eight sections per active volume, thus allowing not only for an accurate measurement of the individual energy loss of the two fission fragments, but also for the determination of their vertical and horizontal positions in the different anode regions, by exploiting drift times and positions of the electrons created by the passing fragments in the counting gas.

The last detector in the set-up was an array of 15 overlapping position-sensitive plastic scintillators, covering an active area of 1 m$^2$. It measured the horizontal position of the fission fragments and, due to its granularity, also their vertical position. It delivered a stop information for a time-of-flight measurement as well.

This set-up determined the nuclear charges of both fission fragments independently through the energy losses in the twin ionisation chamber, in combination with time-of-flight measurements, and gave a resolution of $Z/\Delta Z = 120$. In the following section it will become clear that an excellent charge resolution is crucial to extract fission cross sections after electromagnetic excitation.

A more comprehensive description of the experimental set-up can be found in reference [12] and references therein. The whole set-up is optimised to cope with the limited intensity of the secondary beams, which is caused, on the one hand, by low primary-beam intensities and, on the other hand, by the production cross sections of the projectile-fragmentation reaction.

In our experiment we exploited the possibility to study several secondary beams at the same time. Moreover, our set-up and the rather high kinetic energies of the secondary beams allowed for use of a rather large target thickness and yielded a high detection efficiency due to the forward focusing of the reaction products. Finally, we could distinguish two mechanisms to induce fission of the secondary projectiles, electromagnetic interactions and nuclear collisions, as will be described in the following section. These reactions have large cross sections, of the order of barns, for the isotopes investigated here.

## 3. Data Analysis

The identification procedure of heavy projectile fragments at the fragment separator has been described in detail in previous publications [12, 16]. Here, we give a short summary of the technique used. The comparison of the magnetic rigidity in the first part of the fragment separator with the mass-to-charge ratio in the second part suppressed all ions, which did not maintain the same charge state throughout the whole separator. In the second part of the separator, a time-of-flight measurement, corrected for an angular dependence of the flight path, was used to determine the velocity of the ions. Position measurements at the central and final focal planes were compared to an ion-optical calculation [17]. This identified the nuclear charge by determining the energy loss of the projectile fragments in the intermediate energy degrader. The velocity of the projectile fragments is known from the time-of-flight measurement. Together with the magnetic rigidity the mass can be determined. An example of an identification spectrum is shown in the lower part of Figure 2. The number of counts for a given isotope, as indicated in the lower part of Figure 2, was used for normalizing the fission cross sections. A dead-time correction was not necessary, since the incoming nuclei and the fission products were registered with the same dead time of the data acquisition.

The identified secondary beams have an average energy of 420 $A$ MeV inside the active target. Depending on the impact parameter, two reaction mechanisms contribute to the observed fission events. Interference effects between the two mechanisms can be neglected [18]. If the impact parameter is larger than the sum of the nuclear radii of the secondary projectile and the target, only electromagnetic interactions can contribute to the excitation of the projectile. If the impact parameter is smaller, nuclear interactions, leading to very high excitation energies, will become dominant. Only the first process leads to excitation energies in the vicinity of the fission barrier by mainly populating the electric giant dipole and quadrupole resonances (see ref. [12, 19]). It is necessary to separate the two reaction mechanisms in order to obtain a clear experimental signature.

The two excitation mechanisms show up with different characteristics in the charge-sum spectra, as demonstrated in Figure 4. In order to obtain these spectra, the nuclear charges of the two fission fragments of each fission event were summed up. In contrast to the electromagnetic excitation, nuclear interactions lead to the abrasion of several protons prior to fission with a rather high probability. Therefore, nuclear-induced fission events extend over a large range in the charge-sum spectrum, while electromagnetic-induced fission events form a peak at the nuclear charge of the corresponding secondary projectile.

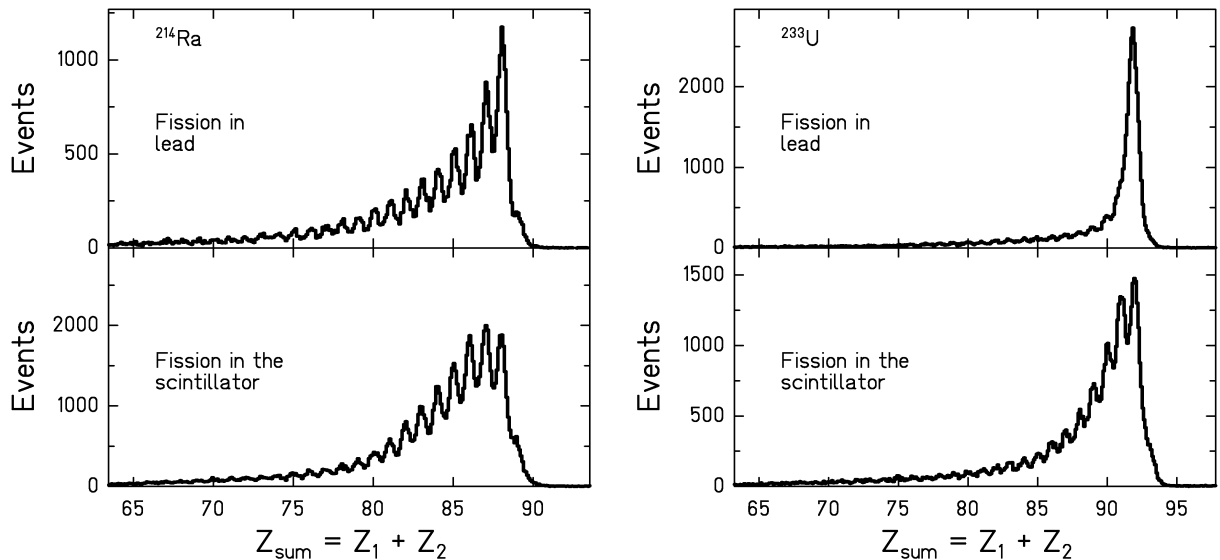

Figure 4: The sum of the nuclear charges of the two fragments from fission of $^{214}$Ra and $^{233}$U in a lead and a scintillator target, respectively.

Here, we use two different procedures to extract the total and the electromagnetic-induced fission cross sections. To obtain the total fission cross sections, the number of fission events in the subdivided scintillation detector was gated on a specific projectile fragment, while also requiring that fission took place inside the lead target. Figure 5 illustrates the identification of fission events on the two-dimensional presentation of the energy-loss signals recorded in both parts of the sub-divided scintillation detector.

There was a finite chance for a reaction in front of the target or inside the counting gas, thereby reducing the number of secondary projectiles that could undergo fission. As another possibility, one of the two fission fragments could undergo a second reaction, reducing its nuclear charge and thus its energy loss. Those events would not be recognized as fission events in the subdivided scintillator. These processes where taken into account by using calculations based on the abrasion-ablation model [9, 20, 21]. The probability of the fission fragments to deexcite via charged-particle emission is small [19] and has been neglected. The total correction factor exceeded in no case the value of 1.6. A correction, taking the detection efficiency of 90 % into account, was also applied.

To obtain fission cross sections after electromagnetic excitation, it was necessary to determine the ratio of the number of fission events after electromagnetic and after nuclear excitation. Since this analysis, which is based on the shape of the charge-sum spectra (see below), requires higher statistical accuracy, a separate measurement was performed, in which the data acquisition recorded only events with a multiplicity of two in the subdivided scintillator, thus reducing the total dead time of the data acquisition. Here only the ratio of the two fission processes was determined, as described in the following. In a first approximation, it can be assumed that the shape, but not the absolute height, of the charge-sum spectrum after nuclear-induced fission is independent of the target material. Thus, events from fission, which took place in the scintillator in front of the lead target, might be interpreted as pure nuclear-induced fission events, as the nuclear charges of its components hydrogen and carbon are low enough to neglect electromagnetic processes. The charge-sum spectrum can be properly normalized and taken as the corresponding charge-sum spectrum for nuclear-induced fission events in the lead target.

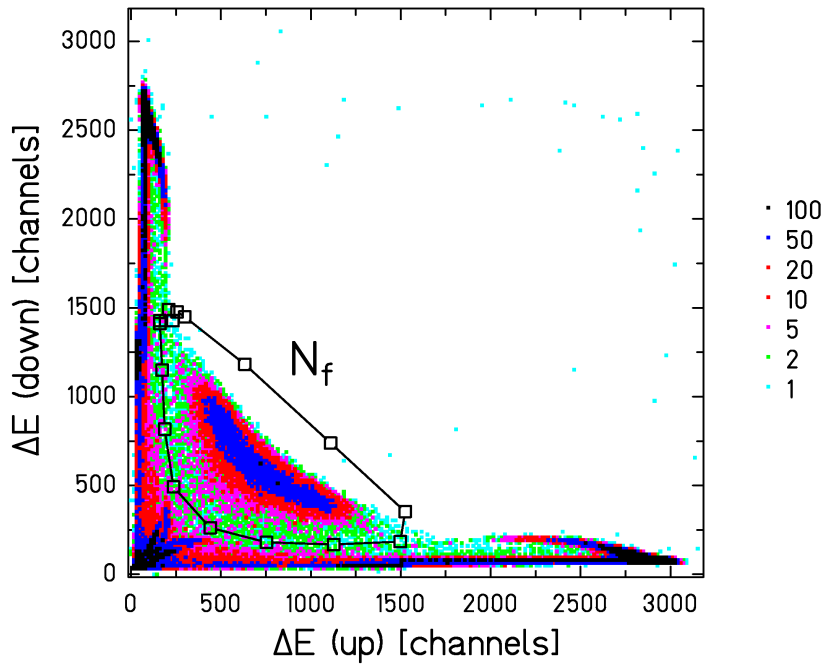

Figure 5: Two-dimensional presentation of the energy-loss signals recorded with the sub-divided scintillation detector. The number of fission events $N_f$ was evaluated from the number of counts inside the polygon window. The scale indicates the number of counts per channel.

To extract electromagnetic-induced fission, all measured charge-sum spectra were fitted with Gaussians in order to determine the number of counts in each peak in the lead ($N_{Pb}(Z_{sum})$) and scintillator target ($N_{Sci}(Z_{sum})$), respectively. The ratio of the two numbers is shown in Figure 6 for $^{214}$Ra and $^{233}$U as examples. Please note the semi-logarithmic representation. In the case of $^{233}$U, a high fission cross section after electromagnetic interactions is expected due to the high fissility of this nucleus. Indeed, the right spectrum of Figure 6 shows a pronounced peak at $Z=92$, indicating fission events without the abrasion of charged particles. As described above, this peak should contain all electromagnetic-induced fission events. As the response of the experimental set-up produces also small non-gaussian tails for the individual charge peaks the spectrum shows an enhancement at $Z=91$ and $Z=93$ as well. A small fraction of the $Z=93$ events results from fission of nuclei with a higher charge than the selected secondary beam, resulting from charge-pickup reactions before fission. The integrated cross section for those reactions has been measured for projectile fragmentation of $^{208}$Pb and found to be less than 30 mb [16]. For charge sums below $Z=85$, the ratio stays constant in a good approximation. This is expected, because nuclear-induced fission events should lead to similar charge-sum spectra, independently of the target. The fact that the ratio never reaches unity reflects the different fission yields according to the target thicknesses used. The decrease for $Z_{sum}>85$ is a result of the hydrogen nuclei in the scintillator target. Spallation reactions of secondary projectiles with hydrogen nuclei lead to significantly lower maximum excitation energies compared with carbon target nuclei [22]. As a result, events with a higher charge-sum contain an increasing fraction of reactions with hydrogen nuclei, and the ratio of the reactions in the two targets (plastic and lead) decreases. The $^{214}$Ra spectrum shows the same characteristics. The peak indicating low-energy fission is weaker than in the $^{233}$U case, but clearly visible. To extract the amount of fission after electromagnetic excitation quantitatively, the decreasing part of the spectrum was extrapolated towards the region of the peak for each isotope individually in contrast to ref. [12], where the same slope was taken for all measured isotopes. This was used to determine the ratio of fission after electromagnetic excitation and after nuclear excitation in the region of the peak. Since the total fission cross section has been determined earlier, this ratio allows the extraction of the cross section for fission events after electromagnetic excitation. A detailed description of this procedure to disentangle the fission events from the two excitation mechanisms can also be found in ref. [12].

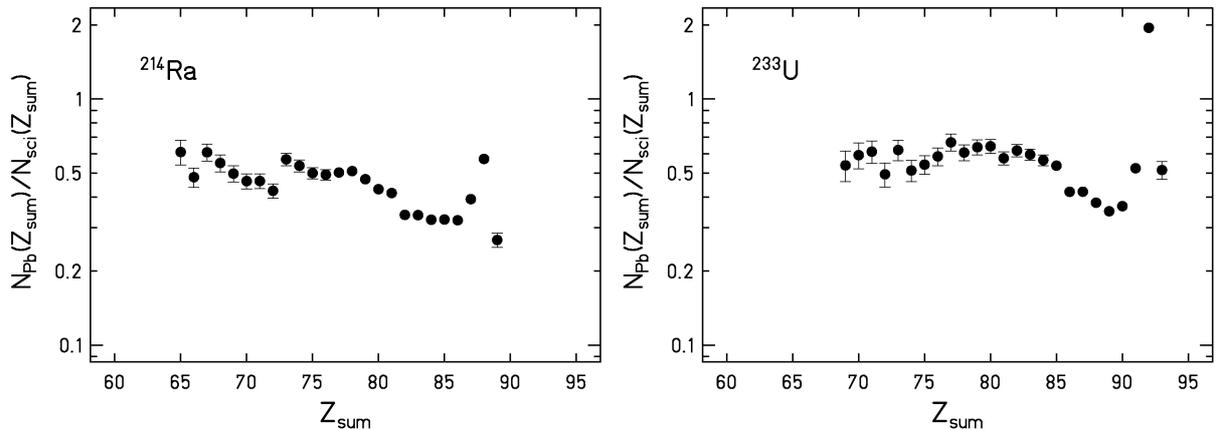

Figure 6: The ratio of counts for individual charge-sums in a lead and a scintillator target using the examples from Figure 4. For details see text.

Due to the analysis procedure which is based on several measurements with different trigger conditions, the nuclear-induced fission cross sections given in tables 1 and 2 are slightly different from the difference of the total and the electromagnetic-induced fission cross sections. These differences are well within the uncertainties of the data.

## 4. Discussion

A few important qualitative conclusions can already be drawn from Figure 7. Here measured charge-sum spectra for a selection of the measured thorium and radium isotopes are shown. The number of counts in the spectra are proportional to the measured total fission cross sections. In accordance with the variation of the fissility parameter $Z^2/A$, the cross section of fission after electromagnetic excitation, visible in the enhanced peaks at $Z_{sum}=90$ and $Z_{sum}=88$, respectively, increases with increasing nuclear charge and decreasing neutron number of the secondary projectiles. The variation in the cross sections of nuclear-induced fission, visible in the left part of the spectra, goes in the same direction, but it is much weaker. Nuclei in the vicinity of the 126-neutron shell do not show any deviation from these global trends. There is no indication for an influence of the shape transition near $N=134$ from spherical to deformed ground-state shapes nor any structure near $N=126$, where the ground-state shell correction attains values up to 6 MeV [23].

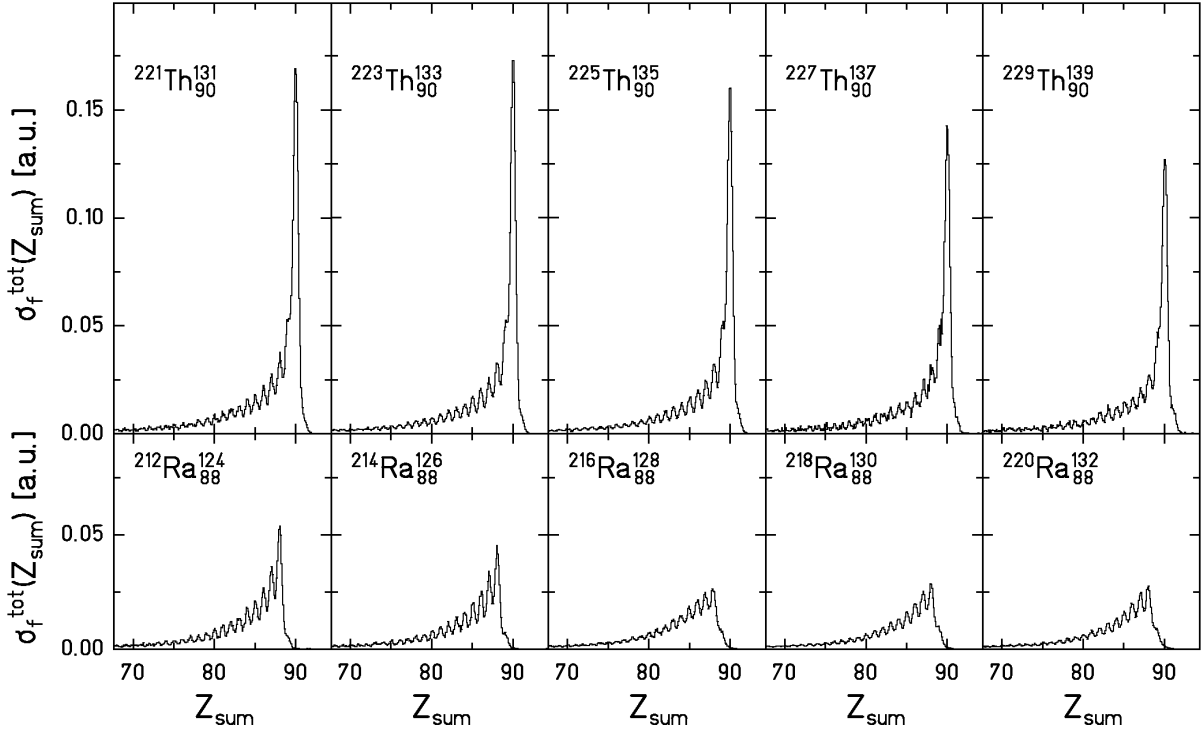

Figure 7: Measured spectra of the sum of the nuclear charges of the two simultaneously produced fission fragments for the isotopes of the elements thorium and radium. The average beam energy in the lead target was 420 $A$ MeV. Each spectrum has been scaled by the corresponding measured total fission cross section. Therefore, they can be compared on a relative scale, although they are not given in absolute units.

A more quantitative analysis can be performed by disentangling nuclear- and electromagnetic-induced fission using the method described in the previous section. Before we discuss the resulting cross sections, we investigate the standard deviations of fission-fragment charge distributions corresponding to fission after nuclear and electromagnetic interaction, shown in Figure 8, because they are a sensitive probe of the excitation energy at fission [24, 25]. These distributions were measured in this work and have been described before in ref. [12]. In the low-mass range ($A_{cn} \leq 221$), where symmetric fission is dominant, the charge distributions of electromagnetic-induced fission show a constant value for all nuclei, including those with very large electromagnetic-fission cross sections like $^{221}$Th and those with quite weak electromagnetic-fission contribution like $^{214}$Ra. Moreover, it is clear that the standard deviation shows a significant difference between nuclear- and electromagnetic-induced fission, also for nuclei near the $N=126$ shell. (The increase for $A_{cn}>221$ is caused by an increasing contribution of asymmetric fission due to the influence of shell effects on the way from saddle to scission.) We expect that electromagnetic fission peaks at an excitation energy of about 11 MeV, while the part of nuclear-induced fission that preserves the number of protons occurs at a mean excitation energy of about 27 MeV [12]. Therefore, the constant value found for the standard deviations of fission-fragment charge distributions after electromagnetic interactions for $A_{cn} \leq 221$ is an indication that the events of electromagnetic-induced fission are correctly identified by the subtraction method, also in the vicinity of the 126-neutron shell. This check gives us confidence that the electromagnetic fission cross sections have correctly been determined in all cases.

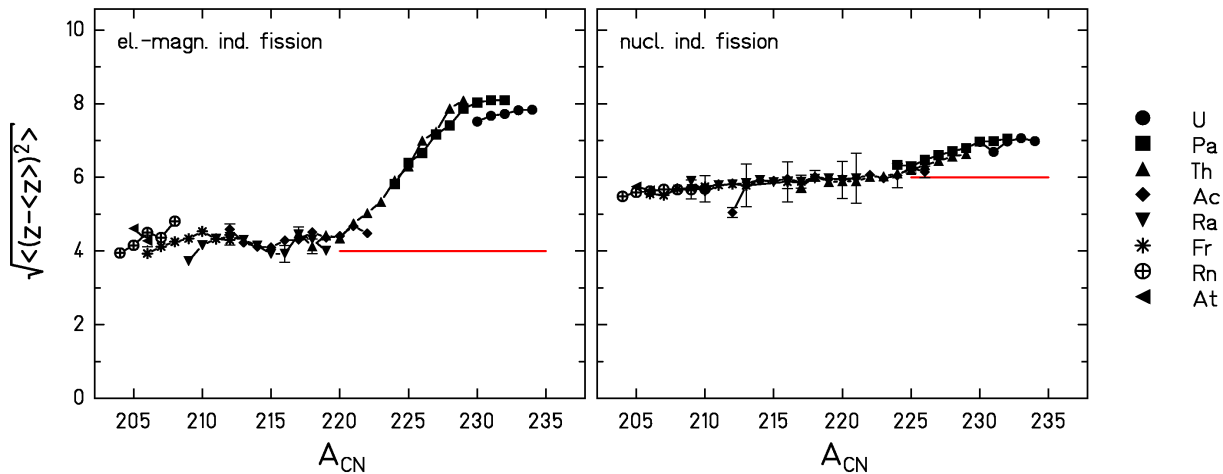

Figure 8: Standard deviation of fission-fragment charge distributions after electromagnetic (left) and nuclear (right) excitation. Also in nuclear-induced fission only those fission events are included where all protons of the secondary projectile are found in the two fission fragments. The method to obtain those distributions has been described in detail in ref. [12].

Figure 9 shows the result of the quantitative determination of the total fission cross sections and of the cross sections for fission after electromagnetic excitation at 420 $A$ MeV in a lead target for the isotopes investigated. We evaluated also the total fission cross sections of secondary projectiles around 300 $A$ MeV in a lead target, measured in a previous experiment [26], which cover some additional nuclei. The data of the two experiments agree well within the given uncertainties. The numerical values are listed in tables 1, 2 and 3. Also in this presentation, the cross sections show a smooth trend, qualitatively explained by the variation of the fissility parameter $Z^2/A$, even in the vicinity of the 126-neutron shell and in the region of the shape transition near $N=134$.

The fission cross sections after nuclear excitations contain interesting information on dissipation in fission. This aspect will be discussed in a forthcoming publication [27]. In the present context, in particular the smooth behaviour of the fission cross section after electromagnetic excitation is a very remarkable finding, because these events originate from excitation energies close to the fission barrier where the structural influence on the fission probability is expected to be strongest. This is illustrated by the calculated energy-differential electromagnetic excitation cross section for $^{214}$Ra in Figure 10. The part of the excitation-energy distribution, which extends beyond the fission barrier, peaks directly at the fission barrier. The height of the fission barrier was estimated as the sum of the prediction of the finite-range liquid-drop model [28] for the macroscopic contribution and the predicted ground-state shell correction [23]. For this magic nucleus directly at the 126-neutron shell, the ground-state shell correction contributes about 50% to the height of the fission barrier.

For all nuclei investigated, which have fission barriers between 5 and 13 MeV, the calculated excitation energies populated by electromagnetic interactions are almost identical [12].

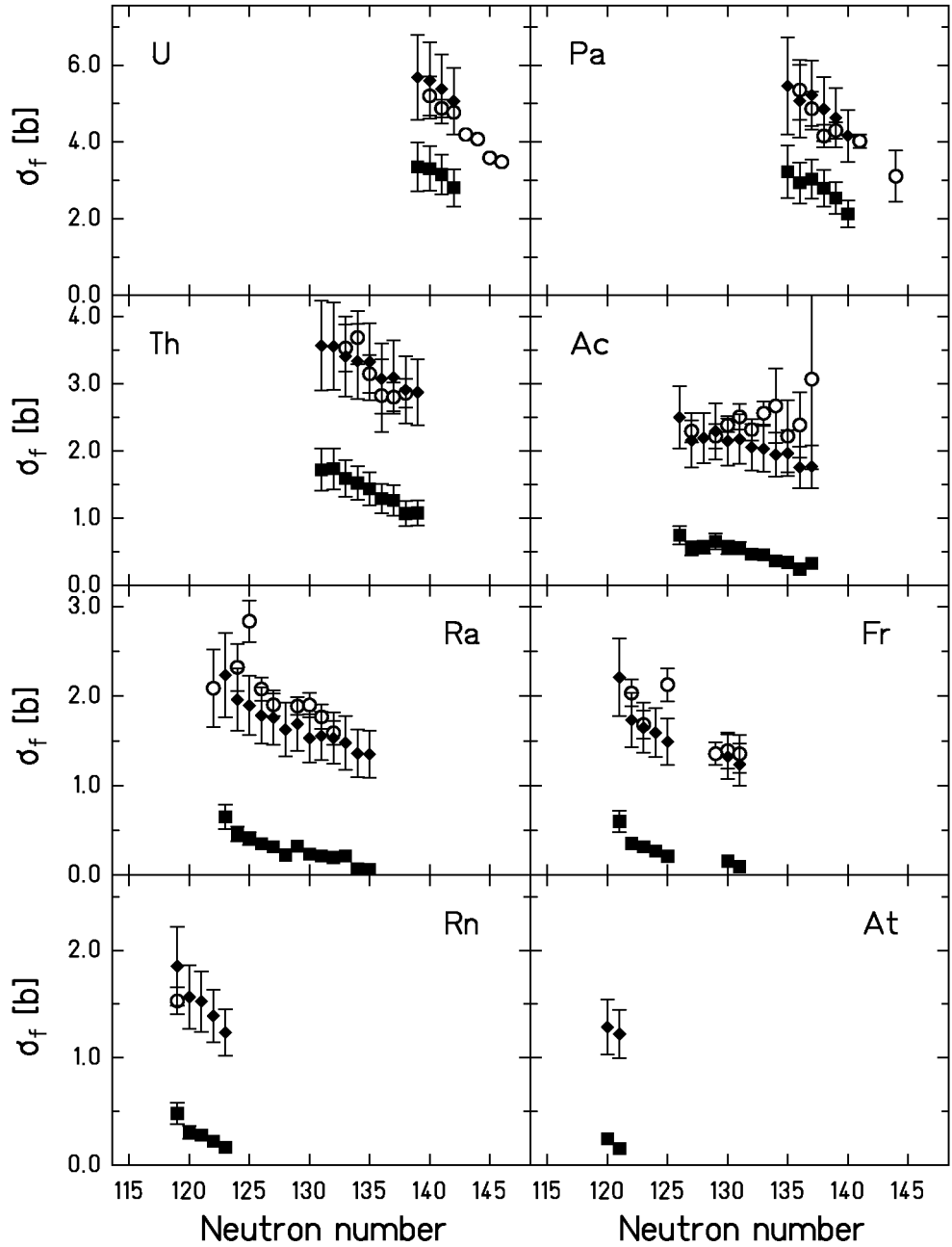

Figure 9: Measured total fission cross sections (full diamonds) and deduced cross sections for fission of secondary projectiles after electromagnetic excitation (full squares) at 420 $A$ MeV in a lead target. The open symbols show the total cross sections of secondary projectiles around 300 $A$ MeV, also in a lead target, obtained in a previous experiment [26].

In order to illustrate the expected influence of the closed 126-neutron shell on the measured fission cross sections after electromagnetic excitation the data are compared to model calculations. In Figure 11 the measured fission cross sections for a number of radium isotopes are shown together with two calculations using the abrasion-ablation model ABRABLA (see reference [9] and references therein). The first calculation (dashed-dotted line) takes the influence of the nuclear shell structure and pairing on the fission barrier height and the level density into account. The calculated fission cross section at $N=126$ underestimates the measured value by about 3 orders of magnitude. The second calculation (solid line) includes also the effect of collective excitations on the level density, leading to an increase of the calculated fission cross sections, which are, however, still far from the measured values. It seems that both calculations predict a stabilisation against fission for nuclei near $N=126$, which is not observed in the data. A possible explanation for these discrepancies might be the specific behaviour of the level density in the energy range considered, in which we represent the energy-dependent influence of shell effects in a global formulation by an exponential function [29] in our calculations.

Figure 11 illustrates the heights of the fission barriers estimated for the nuclei investigated in the present work. The mass model used [23] to extract the ground-state shell effects is in excellent agreement with measured binding energies [30] in the region of the nuclei investigated.

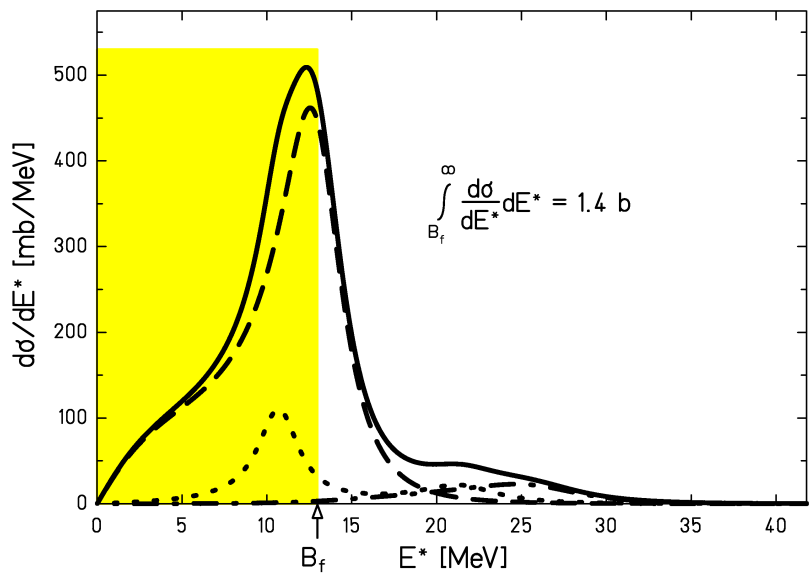

Figure 10: The full line represents the excitation function after electromagnetic interaction for the reaction $^{214}$Ra on $^{208}$Pb at 420 $A$ MeV. The other lines show the different contributions from the one-phonon excitation (dashed line) and the two-phonon excitation (dashed-dotted line) of the giant dipole resonance and the giant quadrupole resonance (dotted line). The fission barrier $B_f$, calculated from values in refs. [28, 23], is indicated. The integrated excitation cross section above the fission barrier is 1.4 b for the case shown.

Finally, we will perform a quantitative analysis of the fission probability of the nuclei investigated. Of course, only a mean value, averaged over the energy range populated by the electromagnetic excitation, can be determined. Unfortunately, this can only be done by applying a model calculation. For each of the systems the fission probability was determined as the ratio of the measured fission cross section that was attributed to electromagnetic excitation and the corresponding part of the calculated electromagnetic excitation cross section which exceeds the fission barrier, see Figure 10 for the case of $^{214}$Ra. Details of the calculation of the differential electromagnetic excitation cross section can be found in reference [12] and in references given therein. The deduced fission probabilities for all systems are depicted in 13. Here, we assume that the contribution of subbarrier fission is small and can be neglected.

The data of the neutron-deficient isotopes of francium, radium and actinium are the most interesting, because these nuclei touch or cross the 126-neutron shell. The fission probabilities deduced with the above-mentioned assumptions show in general a smooth behaviour. A small peak is located directly at the shell on top of the smooth increase with decreasing neutron number. The reason that this peak appears is that only a minor part of the excitation-energy distribution exceeds the fission barrier for these magic nuclei. This is illustrated in Figure 10 for $^{214}$Ra. Our analysis suggests that these magic nuclei tend to fission even more strongly than non-magic nuclei.

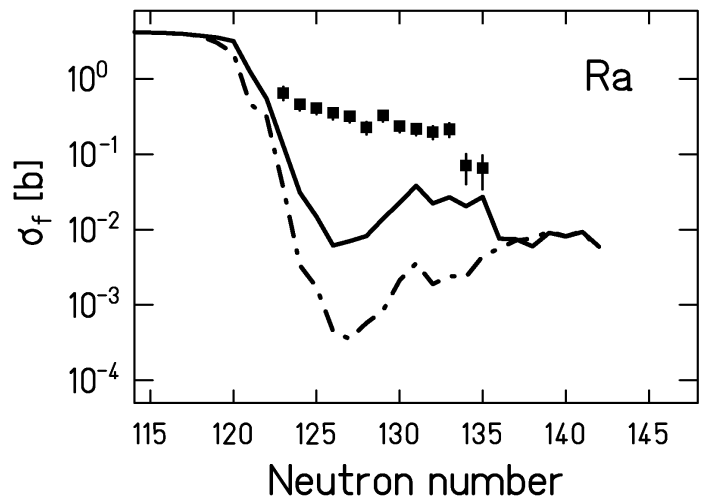

Figure 11: Measured fission cross sections of Ra isotopes after electromagnetic excitation in comparison with model calculations using the code ABRABLA. The dashed-dotted line is a calculation which takes shell and pairing effects in the level density into account. The solid line shows the result of a calculation, which includes collective effects in the level density in addition (for details see text). Please note the logarithmic scale.

Since the fission probability is directly related to the ratio of level densities of the mother nucleus at the fission barrier and of the daughter nucleus after particle, mostly neutron, evaporation, it gives valuable information on nuclear level densities. This aspect has extensively been discussed by A. Junghans et al. [9]. In this context, the present results indicate that shell effects in spherical nuclei do not decrease the fission probability from excited states, even if situated only slightly higher than the fission barrier. These nuclei practically behave like fictive nuclei with liquid-drop binding energies and fission barriers and Fermi-gas level densities. The stabilizing influence resulting from the higher fission barrier seems to be compensated or even over-compensated by the destabilizing effect of the lower collective enhancement in the spherical ground-state shape. Low-lying states in spherical nuclei, belonging to deformed configurations that are not shell stabilized [31], may also enhance the fission probabilities. If these findings can be generalized to other magic nuclei, one expects that one will meet enormous difficulties in the attempts to synthesize spherical super-heavy nuclei near the next double shell closure.

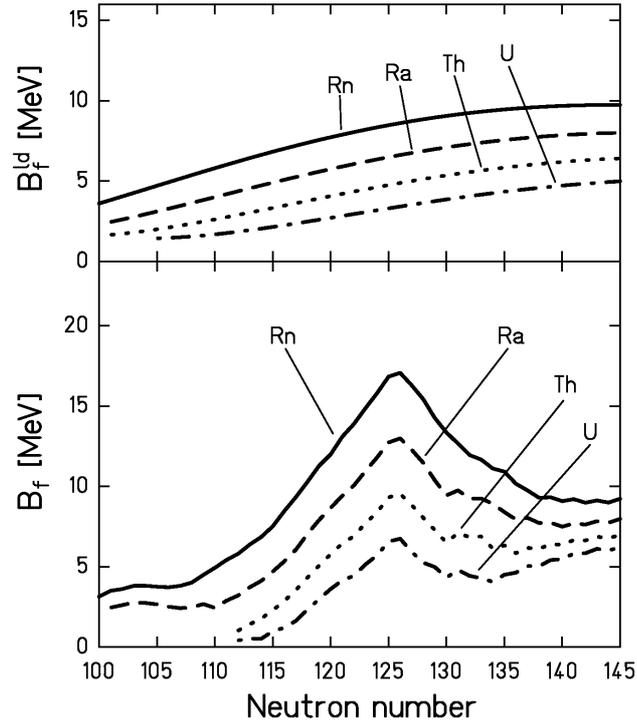

Figure 12: Estimated heights of the fission barriers of the nuclei investigated. The upper part shows the liquid-drop contribution [28], the lower part includes the contribution of the ground-state shell effect [23].

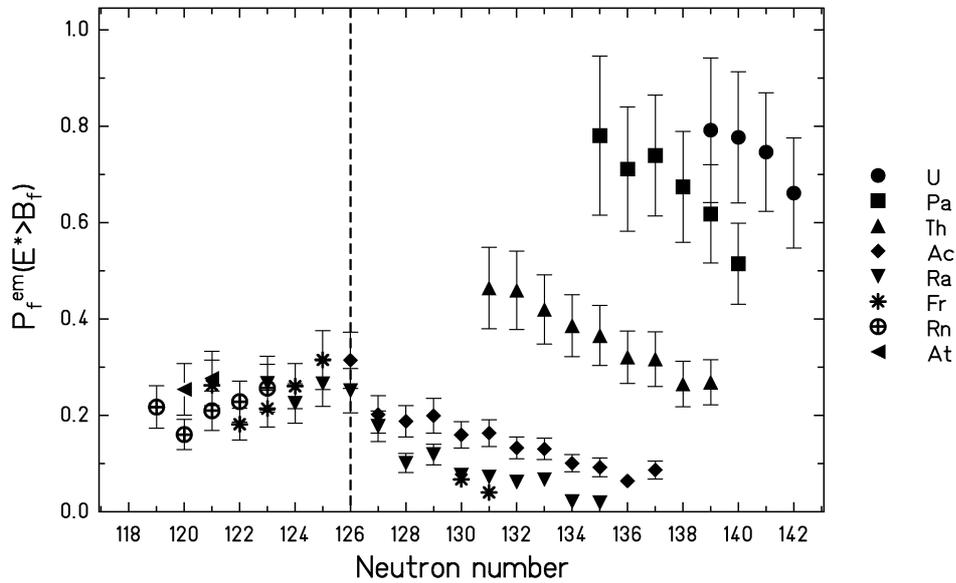

Figure 13: Deduced fission probabilities of secondary projectiles at 420 $A$ MeV in a Pb target due to electromagnetic excitation. The error bars represent the uncertainties of the measured fission cross sections. The assumptions used for the analysis may increase the uncertainties, especially for the lighter systems.

We would like to stress that our finding is not in contradiction to results obtained in refs. [32, 33], where the fission barriers were deduced, including the contribution from shell effects, from fission probabilities at much higher excitation energies, around 100 MeV. In fact, their analysis is based on the assumption that the influence of shell effects on the nuclear level density has completely disappeared, in the sense that it can be expressed by a backshift to the Fermi-gas level density, see also ref. [34].

| Element | Isotope | $\sigma_f^{tot}[b]$ | $\sigma_f^{em}[b]$ | $\sigma_f^{nuc}[b]$ |
|---------|---------|---------------------|---------------------|---------------------|
| U  | 234 | 5.06±0.87 | 2.81±0.49 | 2.26±0.43 |
| U  | 233 | 5.38±0.90 | 3.15±0.52 | 2.23±0.40 |
| U  | 232 | 5.60±1.00 | 3.31±0.58 | 2.29±0.47 |
| U  | 231 | 5.69±1.11 | 3.36±0.64 | 2.33±0.55 |
| Pa | 231 | 4.16±0.68 | 2.13±0.35 | 2.03±0.35 |
| Pa | 230 | 4.63±0.77 | 2.54±0.42 | 2.09±0.37 |
| Pa | 229 | 4.86±0.84 | 2.79±0.48 | 2.07±0.40 |
| Pa | 228 | 5.22±0.90 | 3.04±0.51 | 2.11±0.41 |
| Pa | 227 | 5.07±0.95 | 2.94±0.53 | 2.09±0.46 |
| Pa | 226 | 5.46±1.26 | 3.22±0.68 | 2.24±0.70 |
| Th | 229 | 2.88±0.49 | 1.08±0.19 | 1.80±0.32 |
| Th | 228 | 2.91±0.50 | 1.07±0.19 | 1.85±0.33 |
| Th | 227 | 3.10±0.54 | 1.27±0.23 | 1.84±0.34 |
| Th | 226 | 3.08±0.52 | 1.29±0.22 | 1.74±0.31 |
| Th | 225 | 3.33±0.57 | 1.44±0.24 | 1.82±0.33 |
| Th | 224 | 3.34±0.56 | 1.52±0.25 | 1.85±0.33 |
| Th | 223 | 3.41±0.60 | 1.59±0.27 | 1.83±0.35 |
| Th | 222 | 3.56±0.65 | 1.74±0.31 | 1.90±0.38 |
| Th | 221 | 3.57±0.67 | 1.72±0.31 | 1.84±0.39 |
| Ac | 226 | 1.77±0.32 | 0.33±0.07 | 1.44±0.26 |
| Ac | 225 | 1.75±0.31 | 0.24±0.05 | 1.51±0.27 |
| Ac | 224 | 1.97±0.34 | 0.34±0.07 | 1.63±0.29 |
| Ac | 223 | 1.94±0.33 | 0.37±0.06 | 1.55±0.26 |
| Ac | 222 | 2.03±0.34 | 0.46±0.08 | 1.55±0.26 |
| Ac | 221 | 2.06±0.34 | 0.46±0.08 | 1.58±0.27 |
| Ac | 220 | 2.18±0.37 | 0.55±0.09 | 1.65±0.28 |
| Ac | 219 | 2.15±0.37 | 0.56±0.10 | 1.61±0.28 |
| Ac | 218 | 2.30±0.42 | 0.66±0.12 | 1.64±0.31 |
| Ac | 217 | 2.19±0.37 | 0.57±0.10 | 1.63±0.29 |
| Ac | 216 | 2.16±0.40 | 0.55±0.11 | 1.61±0.31 |
| Ac | 215 | 2.50±0.46 | 0.74±0.14 | 1.70±0.33 |

Table 1: Measured fission cross sections of uranium, protactinium, thorium, and actinium isotopes at 420 *A* MeV in a lead target. Shown are total fission cross sections as well as fission cross sections after electromagnetic and nuclear interaction. The errors include statistical and systematic uncertainties.

| Element | Isotope | $\sigma_f^{tot}[b]$ | $\sigma_f^{em}[b]$ | $\sigma_f^{nuc}[b]$ |
|---|---|---|---|---|
| Ra | 223 | 1.35±0.26 | 0.07±0.03 | 1.29±0.24 |
| Ra | 222 | 1.36±0.27 | 0.07±0.03 | 1.29±0.25 |
| Ra | 221 | 1.48±0.30 | 0.21±0.05 | 1.27±0.26 |
| Ra | 220 | 1.53±0.29 | 0.20±0.04 | 1.34±0.25 |
| Ra | 219 | 1.56±0.27 | 0.22±0.04 | 1.33±0.23 |
| Ra | 218 | 1.53±0.27 | 0.24±0.04 | 1.30±0.23 |
| Ra | 217 | 1.69±0.30 | 0.33±0.06 | 1.37±0.25 |
| Ra | 216 | 1.63±0.30 | 0.23±0.04 | 1.40±0.26 |
| Ra | 215 | 1.76±0.30 | 0.32±0.06 | 1.49±0.26 |
| Ra | 214 | 1.78±0.31 | 0.35±0.06 | 1.43±0.26 |
| Ra | 213 | 1.90±0.33 | 0.41±0.07 | 1.46±0.26 |
| Ra | 212 | 1.96±0.35 | 0.46±0.08 | 1.51±0.27 |
| Ra | 211 | 2.24±0.47 | 0.65±0.14 | 1.59±0.36 |
| Fr | 218 | 1.24±0.24 | 0.10±0.02 | 1.12±0.21 |
| Fr | 217 | 1.32±0.25 | 0.16±0.03 | 1.17±0.22 |
| Fr | 212 | 1.49±0.26 | 0.21±0.04 | 1.28±0.23 |
| Fr | 211 | 1.59±0.27 | 0.27±0.05 | 1.30±0.23 |
| Fr | 210 | 1.65±0.28 | 0.32±0.06 | 1.35±0.23 |
| Fr | 209 | 1.73±0.30 | 0.35±0.06 | 1.52±0.26 |
| Fr | 208 | 2.21±0.44 | 0.60±0.12 | 1.61±0.33 |
| Rn | 209 | 1.23±0.22 | 0.16±0.03 | 0.98±0.17 |
| Rn | 208 | 1.39±0.24 | 0.22±0.04 | 1.18±0.21 |
| Rn | 207 | 1.52±0.28 | 0.28±0.05 | 1.24±0.23 |
| Rn | 206 | 1.56±0.30 | 0.30±0.06 | 1.26±0.24 |
| Rn | 205 | 1.85±0.37 | 0.48±0.10 | 1.37±0.29 |
| At | 206 | 1.22±0.22 | 0.15±0.03 | 1.07±0.20 |
| At | 205 | 1.29±0.26 | 0.25±0.05 | 1.04±0.21 |

Table 2: Continuation of table 1 for isotopes of the elements radium, francium, radon and astatine.

## 5. Achievements and possible developments

The new kind of fission experiment, which is the basis of the present work, relies on the availability of secondary beams of heavy nuclei at relativistic energies. The novel experimental technique developed for these specific conditions has extensively been described in reference [12]. It allows for an important step towards a systematic investigation of the fission properties of nuclei far from stability. Detailed discussions of the new findings concerning the influence of pairing correlations [35] and shell effects [36] on the fission process have been given previously. The present work documents the progress in the investigation of the fission probability from moderately excited states.

We have shown that electromagnetic excitation is a suitable tool for a rather controlled population of excited states close to the fission barrier. The separation of these events from the nuclear background was straightforward for the heaviest elements, thorium, protactinium and uranium, but it became increasingly difficult for the lighter elements. The procedure was checked by the analysis of the standard deviation of the nuclear-charge distributions, but more direct information could be obtained if the masses of the fission fragments were also measured, since nuclear-induced fission is characterized by an increased amount of neutron emission. Such an experiment at the ALADIN and LAND set-up at GSI is in preparation.

The deduced fission probabilities finally rely on a model calculation of the electromagnetic excitation. Indications for some inconsistencies on the high-energy component deduced from recent experimental results [37], might induce some additional uncertainty on the fission probabilities extracted in the present work, but not on the measured fission cross sections. With a controlled excitation of the secondary projectiles by inelastic electron scattering one would precisely determine the excitation energy induced and, in addition, avoid any nuclear background. However, this technique requires a heavy-ion – electron collider ring with a powerful electron spectrometer. Such a facility might be available in the future [38].

| Element | Isotope | $\sigma_f^{tot}[b]$ |
|---------|---------|---------------------|
| U  | 238 | 3.48±0.35 |
| U  | 237 | 3.59±0.37 |
| U  | 236 | 4.08±0.41 |
| U  | 235 | 4.20±0.43 |
| U  | 234 | 4.77±0.49 |
| U  | 233 | 4.88±0.54 |
| U  | 232 | 5.20±0.73 |
| Pa | 235 | 3.11±0.74 |
| Pa | 232 | 4.02±0.44 |
| Pa | 230 | 4.30±0.48 |
| Pa | 229 | 4.16±0.51 |
| Pa | 228 | 4.87±0.66 |
| Pa | 227 | 5.36±0.95 |
| Th | 228 | 2.86±0.36 |
| Th | 227 | 2.80±0.35 |
| Th | 226 | 2.83±0.61 |
| Th | 225 | 3.15±0.42 |
| Th | 224 | 3.69±0.37 |
| Th | 223 | 3.54±0.50 |
| Ac | 226 | 3.07±1.38 |
| Ac | 225 | 2.39±0.54 |
| Ac | 224 | 2.22±0.58 |
| Ac | 223 | 2.67±0.62 |
| Ac | 222 | 2.56±0.31 |
| Ac | 221 | 2.32±0.28 |
| Ac | 220 | 2.51±0.32 |
| Ac | 219 | 2.39±0.26 |
| Ac | 218 | 2.22±0.29 |
| Ac | 216 | 2.30±0.28 |
| Ra | 220 | 1.59±0.21 |
| Ra | 219 | 1.77±0.22 |
| Ra | 218 | 1.90±0.23 |
| Ra | 217 | 1.89±0.21 |
| Ra | 215 | 1.91±0.23 |
| Ra | 214 | 2.08±0.24 |
| Ra | 213 | 2.84±0.37 |
| Ra | 212 | 2.32±0.35 |
| Ra | 210 | 2.09±0.48 |
| Fr | 218 | 1.36±0.25 |
| Fr | 217 | 1.39±0.25 |
| Fr | 216 | 1.36±0.19 |
| Fr | 212 | 2.13±0.28 |
| Fr | 210 | 1.68±0.23 |
| Fr | 209 | 2.04±0.25 |
| Rn | 205 | 1.53±0.20 |

Table 3: Measured total fission cross sections of uranium, protactinium, thorium, actinium, radium, francium, and radon isotopes at 300 *A* MeV in a lead target. The errors include statistical and systematic uncertainties. The experimental set-up has been described in reference [26].

## 6. Conclusion

We conclude that the electromagnetic excitation of secondary projectiles at relativistic energies is the first promising tool to investigate the fission cross sections of exotic nuclei near the 126-neutron shell at low excitation energies. The data obtained up to now with this new method lead to the same conclusions as those deduced from previous experiments performed with conventional methods in which these nuclei could only be produced at appreciably higher excitation energies in heavy-ion fusion and fragmentation reactions. Within the experimental resolution, the data presented here do not show any stabilizing influence of the 126-neutron shell, which would reduce the fission competition in the deexcitation process. This finding might have significant consequences for the production of spherical super-heavy nuclei.

## Acknowledgement

We thank A. Kelić for reading the manuscript. This work has been supported by the GSI Hochschulprogramm and by the Bundesministerium für Bildung, Wissenschaft, Forschung und Technologie under contract number BMBF 06 DA 473.